# Production of Spin-Semiconducting Zigzag Graphene Nanoribbons by Constructing Asymmetric Notch on Graphene Edges

Guang-Yao Song[1], Qing-Hong Yuan[1], Wen-Xin Hu[2], De-Yan Sun[1]

[1]Department of Physics, East China Normal University, No. 500, Dongchuan Road, Shanghai 200241, People's Republic of China

[2]The Computer Center, East China Normal University, No. 500, Dongchuan Road, Shanghai 200241, People's Republic of China

## Abstract

The electronic and magnetic properties of zigzag graphene nanoribbons with asymmetric notches along their edges are investigated by first principle density functional theory calculations. It is found that the electronic and magnetic properties of the asymmetrically-notched graphene nanoribbons are closely related with the depth of notches, but weekly dependent on the length of notches. As the relative depth of notch increases, the energy level of spin-up and spin-down becomes greatly shifted, associated with the gradual increase of magnetic momentum. The asymmetric band shift allows the asymmetrically notched graphene nanoribbons to be a spintronic semiconductor, through which an N- or P-type spin-semiconductor can be obtained by doping B or N atoms.

# I. Introduction

Zigzag graphene nanoribbons (ZGNRs) have received great attention due to its promising potential applications in nanoelectronic and spintronic devices.[1-7] As the two localized edge states are ferromagnetically ordered but antiferromagnetically coupled to each other,[8, 9] ZGNRs were proposed to have many unusual properties such as giant magnetoresistance effect,[10, 11] half metallic,[10, 13] magnetoelectric effect,[14] *etc*. Among which the half-metallic property of ZGNRs has been received a lot of attentions.[15-18] Son *et. al*. predicted that the semiconductor ZGNR could be changed as half-metallicity under in-plane transverse electric fields across the zigzag-shaped edges.[12] Hod *et. al*.[19] reported that the onset electron field required to induce half-metallic behavior of ZGNRs could be lowered by edge oxidation. Kan *et. al*.[13] further showed that without external electron field ZGNRs could also be half-metals by functionalizing the ribbon with -$NO_2$ groups on one edge and -$CH_3$ groups at the other. The aforementioned studies further promote applications of ZGNRs in nanoelectronics.

Phusically, the electronic property of ZGNR is mainly attributed to its localized $\pi$ and $\pi^*$ states on the edge, thus the edge engineering has been proposed as a route to tailor various properties of ZGNR. Nowadays, engineering of GNRs' edge with designed structure has become feasible with the experimental advances in fabricating and patterning of graphene nanoribbons.[20] Based on this, theoretical investigations on ZGNRs with engineered edges have been strongly motivated. Wu *et. al.* investigated the electronic and magnetic properties of hydrogen passivated ZGNRs (H-ZGNR) with symmetric protruded steps on both edges. They found that the electronic and magnetic properties were closely related with the length of the protruded step and the distance of two adjacent steps along the ribbon edge.[21] However, a ZGNR with symmetric edges is not an efficient spin injector due to the opposite magnetization between two edges. In order to obtain net spin injection, this symmetry must be broken. One possible way to achieve this is to break the magnetic symmetry of the ZGNR edges.[4,20-21] Theoretical calculations did show an interesting spin behavior of ZGNR when a V-shaped notch was produced on the single edge of ribbon.[4, 22] Although previous efforts have implied that an asymmetric notch on edges could realize net spin injection in a ZGNR, what kind of shape of notches is relevant and important is less investigated. In fact, quite a few open questions need to be addressed before it can be easily achieved in experiments and technology. For examples, how the relative shift between spin-up and spin-down bands is correlated with the shape and depth of the notches is not clear. And the major factor affecting the spin band shift is also less investigated. What's more, for realistic applications, an N- or P-doping spin-semiconductor based on the spin behavior of engineered ZGNRs requires more deep studies.

In this work, based on first principle density functional calculations, we studied the spin electronic property of ZGNRs with asymmetric notches on its edges (ASN-ZGNR). It was found that the spin-up and spin-down states could be separated from each other through introducing asymmetric notch on the single edge of ZGNR. With the increase of notch depth, the valence and conduct bands of one spin state shift greatly from the other one, associated with the gradual increase of magnetic momentum. Similar behavior was found on the hydrogen passivated ASN-ZGNR. The asymmetric band shift allows the ASN-ZGNR to be a spintronic semiconductor. By doping boron or nitrogen atoms in the ASN-ZGNRs, an N- or P-type spin-semiconductor device can be obtained. Considering experimental advances have been achieved on the modification of graphene nanoribbons with notched edges,[23] the asymmetry of electronic band shift for the different spins found in our research paved a way for experimentalist to explore new spin injectors or detectors in graphene spintronics by using ZGNR.

## II. Computational Details

ZGNRs with symmetric and asymmetric notches are considered in our study, the detailed computational model can be found in Fig. 1. For ZGNR with symmetrically notched edges (SN-ZGNR), the notches with length of $l$ and depth of $d/2$ are symmetrically distributed on both edges (Fig. 1a). For ASN-ZGNR (Fig. 1b), the notch is produced only on one edge of the ZGNR, with the length of $l$ and depth of $d$, respectively. The length and the width of pristine ZGNRs are defined as $L$ and $W$ respectively. The periodic boundary condition is applied to the direction parallel to edges, namely length direction. The notched structure is produced by cutting the edge of ZGNR along a zigzag orientation which is 60 degree deviated from the zigzag edge. Each moving of C atoms from the ZGNR's edge make the notch appears as V-shape or near V-shape. For all structures, the length of ZGNR is fixed as 2.95 nm in the super cell. To exclude the interaction between neighboring notches, the distance between neighboring notches is as far as 1.72 nm.

All structures and electronic/magnetic properties are investigated using a density functional theory (DFT) method as implemented in the Vienna Ab initio Simulation Package (VASP).[24] The ion-electron interactions are treated with the projected augmented wave (PAW) pseudopotentials[25] using the Perdew-Burke-Ernzerhof (PBE) exchange-correlation functional [26] for the spin-polarized generalized gradient correction. Symmetry-unrestricted optimizations of both geometry and spin are performed to relax the geometries until the energy change is less than $10^{-4}$ eV. The Brillouin zone is sampled by 1×1×1 along the periodic direction using the Monkhorst-Pack scheme.

## III. Results and discussion

The electronic property of SN-ZGNR was firstly explored. Fig. 2a shows the band structure and density of state (DOS) for the SN-ZGNR. It can be seen that the SN-ZGNR has an antiferromagnetic (AFM) ground state in which the two degenerated spin electronic states near the Fermi level are ferromagnetic ordered at the two edges. The flat band near the Fermi level is originated from the edge states of the SN-ZGNR (the valence band maximum as shown in Fig. 2b), which is similar to a pure ZGNR. The magnetic momentums are symmetrically distributed on both edges of SN-ZGNR (Fig. 2c), leading to a zero total momentums. Hence, it can be concluded that the band structure of SN-ZGNR is quite similar to that of pristine ZGNR despite that both the energy gap and energy bands can be slightly modified. Therefore, the SN-ZGNR is not suitable for the spintronic applications because of the symmetric spin-up and spin-down states.

ASN-ZGNR shows quite different spin electronic behavior around the Fermi level. As shown in Fig. 2d, an opposite energy band shift of the spin polarized edge states is observed; *i.e.,* the spin polarized DOS for spin-up (spin-down) states near the Fermi level are shifted downwards (upwards). Meanwhile, the energy gaps for each spin state in ASN-ZGNR are slightly increased in comparison with that in SN-ZGNR. For example, the energy gaps of spin-up and spin-down states in ASN-ZGNR are ~0.8 eV (Fig. 2d), while they are 0.6 eV in SN-ZGNR (Fig. 2a). Such an increase of energy gaps could be attributed to the more localized valence band maximum (VBM) of spin-up state and conducting band minimum (CBM) of spin-down state in ASN-ZGNR, as shown in Fig. 2e. Given that there is no transverse electric field applied in the concerned system, the opposite energy-level shift for the spin-polarized states could be realized by the asymmetric notch.

In addition to the electronic property, the magnetic property between SN-ZGNR and ASN-ZGNR

is quite different. As can be seen from Fig. 2c and 2f, the SN-ZGNR shows no magnetism while ASN-ZGNR has a magnetic momentum of 16 μB. The different magnetism between SN- and ASN-ZGNR is attributed to the different number of sublattice atoms. It was reported that the magnetic momentum was proportional to the number difference of atoms in sublattice: $|n_A － n_B|$, where $n_A$ ($n_B$) is the number of atoms on the A (B) sublattice of the GNR.[27] For SN-ZGNR, $n_A$ always equals to $n_B$ and thus there is no magnetism in SN-ZGNR. While for the ASN-ZGNR shown in Fig. 2f, there are 210 carbon atoms in the unit cell, among which 99 atoms belong to the A site and 111 atoms belong to the B site, leading to $|n_A － n_B|=12$.

The above results demonstrated that the electronic structure of spin-up and spin-down could be manipulated separately by the asymmetric notch. While how large the band shift can be achieved and what is the major factor affecting the band shift is not clear. To answer these questions, we explored the relationship between the band shift of the different spin states and the variation of notches. We firstly investigate an ASN-ZGNR with fixed ribbon width (*W*) of 1.988 nm, and explored the variation of the electronic property with the change of the notch depth (*d*). It is found that, the electronic structure strongly depends on the depth of notch, while it is less relevant to the length of notches. Fig. 3a shows the spin energy gap versus to the notch depth (*d*). It can be seen that the energy gaps of both spin states increase gradually with the increase of *d*, and the difference between the two spin states become more and more significant. When *d* is less than 0.64 nm, the energy gaps of each spin states have comparable magnitude. As *d* varies from 0.64 to 0.85 nm, the energy gaps of each spin state experience an obvious jump. When *d* is larger than 0.85 nm, the energy gaps of spin-up states keep increasing, but the energy gaps of spin-down states almost keep unchanged. In accordance with the gradual increase of energy gap, the magnetic momentums of ANS-ZGNRs rise with the increase of *d*, as shown in the inset of Fig. 3a. This is understandable since the magnetic momentums of ASN-ZGNRs are proportional to the difference between $n_A$ and $n_B$. The deeper notch (or larger *d*) corresponds to more removing of type A atoms from the edge and thus bigger $|n_A － n_B|$.

We notice that the electronic property of ASN-ZGNR experienced a remarkable change when the *d* varies from ~0.64 to ~0.85 nm. Such a change is represented by two aspects. On the one hand, the energy gaps of both spin states are suddenly increased. On the other hand, the energy bands around the Fermi levels of spin-up and spin-down state are largely shifted (Fig. 3a). To understand such changes, we calculated the DOS of structure A, B, C and D. As shown in Fig. 3b, the Fermi levels of the spin-up and spin-down states are gradually shifted when the structure changes from A to D. For structure A and B with shallow notches, the two spin states are much similar and the energy gaps of the spin-up and spin-down states have no much difference. While for structure C and D, the electronic band of spin-up and spin-down states are significantly shifted. Meanwhile, the energy gaps for the two spin states in structure C and D are notably increased, e.g. the spin energy gaps for structure D is ~0.2 eV larger than that of structure A and B.

The sudden increase of energy gap from structure B to C could be attributed to the increased localization of edge states. To understand this point, we compared the band structures, partial charge density distributions, as well as VBMs and CBMs of the spin-up and spin-down states for structure B and C, as shown in Fig. 3c and 3d. We can see that the band structure of spin-up and spin-down states in B are almost the same. And the VBMs and CBMs for spin-up and spin-down states are only slightly delocalized. However, for structure C, the energy bands near the Fermi level are flatter, and partial charge density distributions of VBMs and CBMs are more localized.

Based on the above results, we can conclude that the opposite energy-level shift for the spin polarized edge states is closely related with the depth of the notch and the shift becomes significant when $d$ is larger than 0.8 nm. Interestingly, the length of the notch has little effect on the electronic property of the ASN-ZGNR if the length is larger than certain value (here, 1nm). With the increase of notch length ($l$), the band gap of the ASN-ZGNR keeps almost the same, as shown in Fig. 4.

Interestingly, the band gap seems mainly depend on the relative depth ($d/W$), the ratio of notch depth to ribbon width. We have made a few additional calculations for systems with different $d$ and $W$ but similar $d/W$. We have found a similar electronic properties if $d/W$ is closed. For examples, for ASN-ZGNR with absolute notch depth of 0.426 nm and width of 1.988 nm, the band gap difference between spin-up and spin-down states is as small as 0.04 eV. When the absolute notch depth of ASN-ZGNR is increased to 0.639 nm but the $d/W$ keeps almost the same ($d/W$= 0.22), the energy gap difference between the spin-up and spin-down states is almost the same as the former ASN-ZGNR. When the absolute notch depth is kept as 0.639 nm but the $d/W$ is increased to 0.56, the gap difference between the spin-up and spin-down states is increased to 0.12 eV. This demonstrates that the relative notch depth of ASN-ZGNR plays an important role in the separation of spin electronic states.

The above calculations are carried out for ZGNR without any passivation on its edge. In real experiments, the edges of ZGNRs could be passivated by some inorganic groups such as H, OH *etc*. To explore the effect of edge passivation on the spin electronic property of ASN-ZGNR, we further studied the electronic properties of hydrogen saturated ASN-ZGNR (H-ASN-ZGNR). As shown in Fig. 5, different from ASN-ZGNR, H-ASN-ZGNRs have almost degenerated energy gaps for the spin-up and spin-down states. However, the opposite band shift of spin-up and spin-down states keeps unchanged. Therefore, it can be concluded that the energy-shift behavior in the ASN-ZGNR is maintained after the hydrogen passivation.

The energy band shift of the spin-up and spin-down states potentially make the H-ASN-ZGNR a promising spintronic device. From the DOS of structure B shown in Fig. 5b, we can see the system will become an n-type semiconductor with spin-down electrons when electrons are doped, while the system will be a p-type semiconductor with spin-up holes if holes are doped. Existing proposals to achieve the energy band shift of ZGNR is to apply transverse electric fields on ZGNR,[12] which may finally leads to half-metal or spin-semiconductor GNR. Here we propose that, with a V-shaped notch on single edge of ZGNRs, it is also possible to break the spin symmetry, and leading to an asymmetric spin properties that may have application in spintronics.[4, 28]

The last question we should answer is, after the doping was made, whether or not the n- or p-type semiconductor could be achieved as expected. To answer this question, we have studied the electronic properties of H-ASN-ZGNR doped by nitrogen and boron atoms. Fig. 6a shows the DOS for the nitrogen doped, boron doped and undoped H-ASN-ZGNR. Since the nitrogen atom has five valence electrons, doping one nitrogen atom in the system will lead to an n-type semiconductor, and the doped electron occupies the spin-up state, as shown in Fig. 6b. One can see that, the VBM and CBM of spin-down state are little affected by the doping. Similarly, a p-type semiconductor in which all holes have the same spin can be achieved by born doping. The doped holes of boron-doped H-ASN-ZGNR occupies the spin-down state. Thus a typical spin-semiconductor is obtained by doping the H-ASN-ZGNR with nitrogen or boron atoms.

## CONCLUSION

The separated energy band shift of spin-up and spin-down states can be achieved by introducing asymmetric notch in a single edge of ZGNRs. The amplitude of energy shift between spin-up and spin-down states is majorly affected by the depth of notches. Such a behavior is rarely affected by the hydrogen passivation of ASN-ZGNRs' edges. Through nitrogen or boron doping in the hydrogen-passivated ASN-ZGNR, a spin-semiconductor can be obtained, which may have potential applications in future spintronic devices.

## ACKNOWLEDGMENTS

This research is supported by the Natural Science Foundation of China (Grant No.11174079), National Basic Research Program of China (973, Grant No. 2012CB921401), Shuguang Program of Shanghai Education Committee. The computation is performed in the Supercomputer Center of ECNU.

**Fig. 1:** (Color online) Computational model for ZGNR with (a) symmetric and (b) asymmetric notches on the edge. Here *l* and *d* (*d/2*) refer the length and depth of notches respectively. The length and the width of pristine ZGNRs are labeled as L and W respectively.

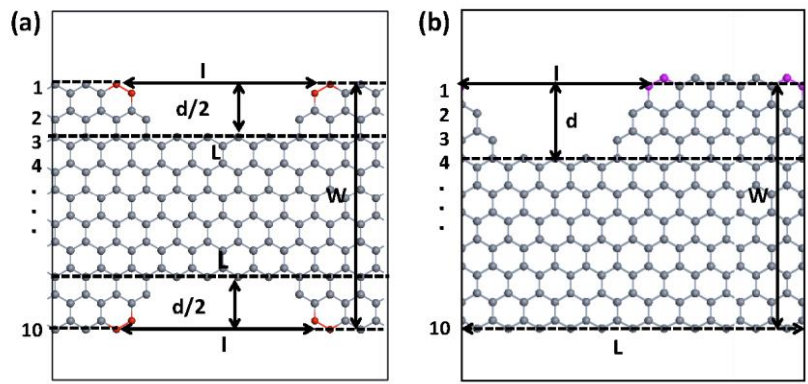

**Fig. 2**: (Color online) The electronic and magnetic properties for the symmetrically (a-c) and asymmetrically (d-f) notched ZGNRs. Here the symmetric notch is produced by moving two rows of V-shaped C atoms from both edges of ZGNR, while the asymmetric notch is produced by moving four rows of V-shaped C atoms from the single edge of ZGNR. Band structure and spin polarized DOS of (a) SN-ZGNR and (d) ASN-ZGNR; Band-decomposed charge densities of VBM and CBM of (b) SN-ZGNR and (e) ASN-ZGNR; The magnetic moment distributions (red represents spin up density and blue represents spin down density) of (c) SN-ZGNR and (f) ASN-ZGNR.

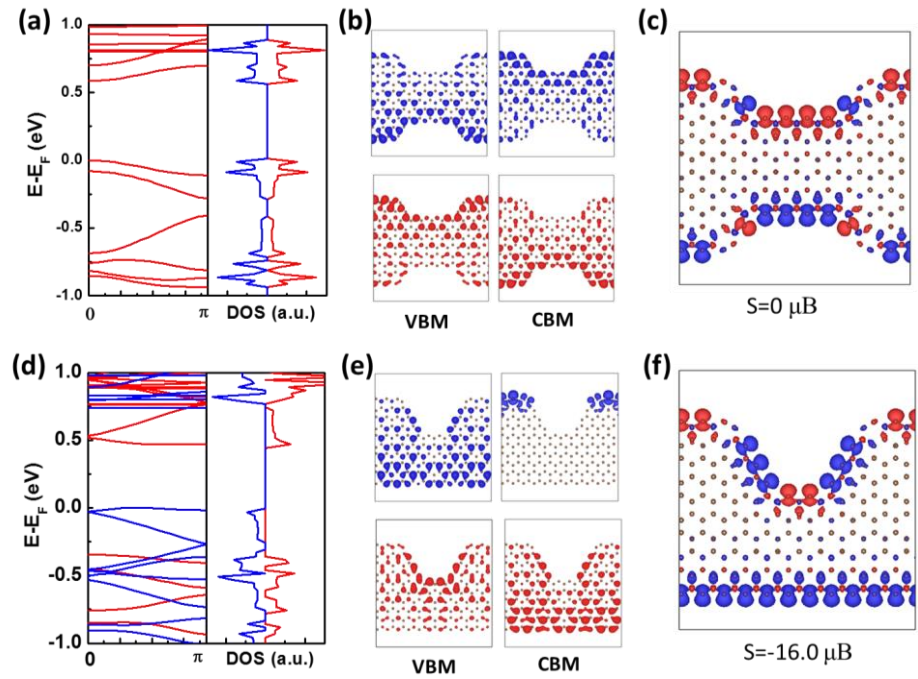

**Fig. 3:** (Color online) (a) Band gap of ASN-ZGNRs with fixed W and variable *d*; (b) DOS of ASN-ZGNRs with different *d* as shown in (a); Band structure and band-decomposed charge densities of VBM and CBM for structure B (c) and C (d). The capital letters (A, B, C, D) refer to the system labeled in (a).

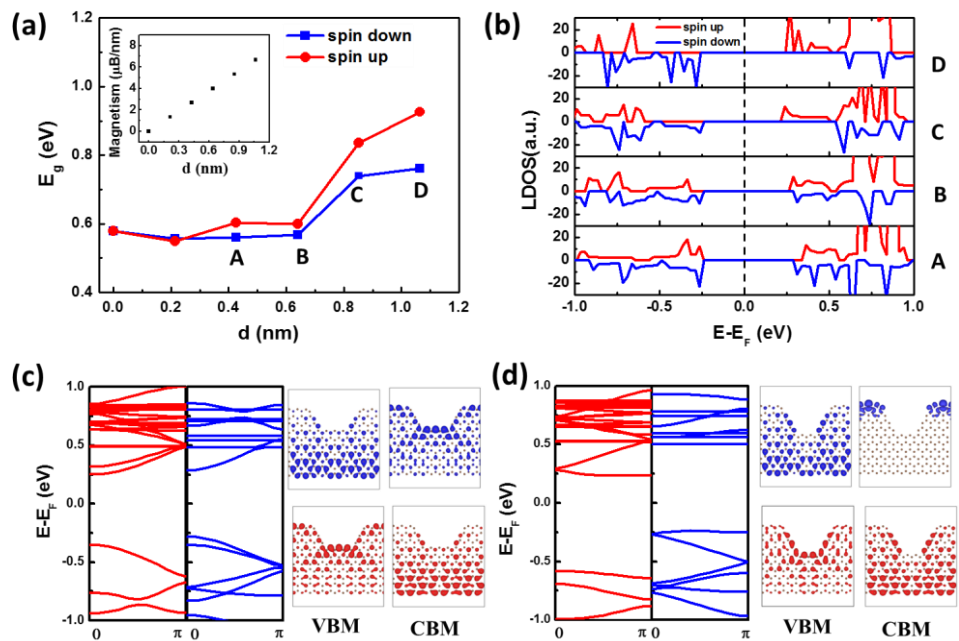

**Fig. 4:** (Color online) The energy gaps of ASN-ZGNR with the variation of notch length. If the length is larger than 1nm, the band gap keeps almost unchanged.

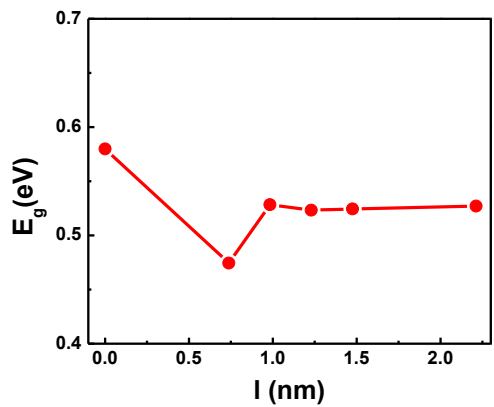

**Fig. 5:** (Color online) (a) Energy band gaps of the hydrogen passivated ZGNRs with asymmetric notches as a function of d/W; (b) DOS of H-ZGNRs with different notch depth as shown in (a).

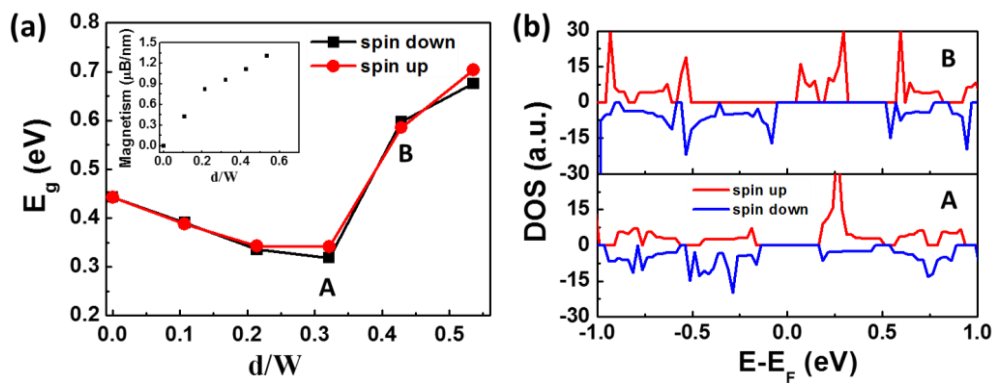

**Fig. 6:** (Color online) (a) DOS for the nitrogen doped, boron doped and undoped H-ASN-ZGNR; (b) Band-decomposed charge densities of VBM and CBM for the nitrogen doped (upper panel), boron doped (middle panel) and undoped (lower panel) H-ASN-ZGNR

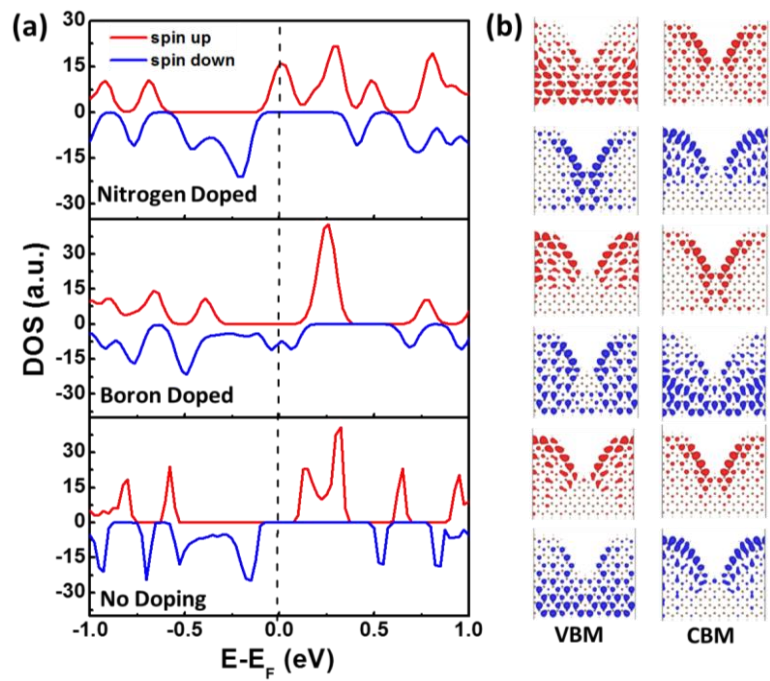